%% file: main.tex
\documentclass[sigconf, authorversion]{acmart}

\AtBeginDocument{%
  }

\copyrightyear{2025}
\acmYear{2025}
\setcopyright{rightsretained}
\acmConference[CHI EA '25]{Extended Abstracts of the CHI Conference on Human Factors in Computing Systems}{April 26-May 1, 2025}{Yokohama, Japan}
\acmBooktitle{Extended Abstracts of the CHI Conference on Human Factors in Computing Systems (CHI EA '25), April 26-May 1, 2025, Yokohama, Japan}\acmDOI{10.1145/3706599.3720167}
\acmISBN{979-8-4007-1395-8/2025/04}

\usepackage{enumitem}
\usepackage{url}

\usepackage{CJK}

\begin{document}
\begin{CJK}{UTF8}{gbsn}

\title[Designing Human-AI System for Legal Research]{
Designing Human-AI System for Legal Research: \\A Case Study of Precedent Search in Chinese Law}

\author{Jiarui Guan}
\authornote{Both authors contributed equally to this research.}
\affiliation{%
  \institution{Tongji University}
  \city{Shanghai}
  \country{China}}
\email{2251196@tongji.edu.cn}

\author{Ruishi Zou}
\authornotemark[1]
\affiliation{%
  \institution{Tongji University}
  \city{Shanghai}
  \country{China}}
\email{zouruishi@tongji.edu.cn}

\author{Jiajun Zhang}
\affiliation{%
  \institution{China University of Political Science and Law}
  \city{Beijing}
  \country{China}}
\email{zhjj_e_email@126.com}

\author{Kimpan Xin}
\affiliation{%
  \institution{Tongji University}
  \city{Shanghai}
  \country{China}}
\email{2250283@tongji.edu.cn}

\author{Bingsu He}
\affiliation{%
  \institution{Tongji University}
  \city{Shanghai}
  \country{China}}
\email{hbsssss@tongji.edu.cn}

\author{Zhuhe Zhang}
\affiliation{%
  \institution{Tongji University}
  \city{Shanghai}
  \country{China}}
\email{Tinuvile@tongji.edu.cn}

\author{Chen Ye}
\authornote{Chen Ye is the corresponding author.}
\affiliation{%
  \institution{Tongji University}
  \city{Shanghai}
  \country{China}}
\email{yechen@tongji.edu.cn}

\begin{abstract}
Recent advancements in AI technology have seen researchers and industry professionals actively exploring the application of AI tools in legal workflows. Despite this prevailing trend, legal practitioners found that AI tools had limited effectiveness in supporting everyday tasks, which can be partly attributed to their design. Typically, AI legal tools only offer end-to-end interaction: practitioners can only manipulate the input and output but have no control over the intermediate steps, raising concerns about AI tools' performance and ethical use. To design an effective AI legal tool, as a first step, we explore users' needs with one specific use case: precedent search. Through a qualitative study with five legal practitioners, we uncovered the precedent search workflow, the challenges they face using current systems, and their concerns and expectations regarding AI tools. We conclude our exploration with an initial prototype to reflect the design implications derived from our findings.
\end{abstract}

\ccsdesc[500]{Human-centered computing~Human computer interaction (HCI)}
\ccsdesc[500]{Applied computing~Law}

\keywords{Human-AI collaboration, Law, Precedent search}

\maketitle

\input{sections/1-Intro}
\input{sections/2-RelatedWork-short}

\input{sections/3-Method}

\input{sections/4-Finding}

\input{sections/5-Discussion}

\input{sections/8-Conclusion}

\begin{acks}
This work was supported in part by the Shanghai Student Innovation Training Program \#S202410247229. We want to thank all participants for their valuable insights. Additionally, we extend our thanks to the anonymous reviewers for their constructive feedback.
\end{acks}

\balance



\appendix

\input{sections/9-appendix}

\end{CJK}
\end{document}

%% file: sections/1-Intro.tex
\section{Introduction}
\label{sec:intro}

Recent advancements in artificial intelligence (AI) have attracted growing interest from academia and industry in using AI tools to support legal practices~\cite{chen2024a}\footnote{In this paper, we use the term ``AI legal tools'' to refer to a broad spectrum of AI-infused systems~\cite{amershi2019guidelines} that are either designed specifically for legal application or generic tools (e.g., chatbots) that can be repurposed for legal tasks.}. According to a 2024 survey, 76\% of legal departments and 68\% of law firms use AI legal tools at least once per week~\cite{WoltersKluwer2024report}. 
Despite the high adoption figures, practitioners found those tools less than satisfactory. Legal practitioners are among the least benefited by AI tools: research has found that the legal domain is among the lowest compared with other domains in productivity gain under AI adoption~\cite{jaffe2024generative}.

Among possible reasons for the contrast between the enthusiasm and the effectiveness of AI legal tools, one critical concern derives from the AI algorithms' design. Today's AIs are mostly unexplainable, raising concerns regarding fairness, trustworthiness, and ethics~\cite{mehrabi2021survey, caton2024fairness} using AIs in the high-stake, ``critical societal domain''~\cite{chen2024a} of law. 
Specifically, emerging AI legal tools (e.g., \cite{metalaw, pkulaw-ai}) normally apply an end-to-end approach (i.e., users only control input/output) in assisting legal tasks. Such an approach left no space for users to intervene in the intermediate processes of AI algorithms, raising doubts from both ethical and practical grounds using AI-generated results. In extreme cases, the approach might even lure practitioners into overtrusting AI results, posing significant risks for themselves and their clients (e.g., ~\cite{weiser2023heres}).

One prevalent way to resolve the conflict between human control and AI is through designing systems following the human-AI collaboration paradigm~\cite{wang2020humanai}. Through careful human-AI interaction design~\cite{horvitz1999principles, amershi2019guidelines}, users can maintain certain agencies using the system while benefiting from AI assistance. Applications in other critical societal domains (e.g., healthcare) have demonstrated that systems following the human-AI collaboration paradigm can reap benefits in real-world high-stakes scenarios (e.g.,~\cite{zhang2024rethinking}). %

Despite the potential of human-AI collaboration, works at the intersection of computation and law are nascent in exploring this paradigm. Although some claimed to have designed a human-AI system~\cite{han2024legalasst}, we need further work through the lens of legal experts to understand the legal workflow--a first step in better aligning AI legal tools with user needs. Informed by this gap, we raised our research question: \textbf{\textit{How should we design a human-AI legal research tool that enables legal practitioners while mitigating the risks of end-to-end AI tools?}}

As a first step, we focused on one use case: \textbf{\textit{precedent search}}. We chose this use case because 1) precedent search is an essential task for legal practitioners across the world regardless of the legal system~\cite{alexander1989constrained}, 2) precedent search is an established ground for algorithmic application (i.e., with traditional search), presenting a more immediate ground for introducing AI support, and 3) precedent search is instrumental but tedious and labor-intensive, an ideal candidate scenario where AI legal tools could benefit legal practitioners. 

Additionally, we scope our study within the context of the Chinese legal system. While the weight of precedents varies across legal systems, we argue that the Chinese system, which blends civil and common law practices, presents an ideal arena for our case study. This hybrid nature allows us to derive insights potentially generalizable to civil and common law systems. With the precedent search use case and the Chinese law context, we aimed to answer:
\begin{enumerate}
    \item[\textbf{RQ1}] What is the legal practitioner's workflow in precedent search?
    \item[\textbf{RQ2}] What are the challenges in today's legal tools that impose friction in precedent search?
    \item[\textbf{RQ3}] What are the concerns and expectations of legal practitioners in AI legal tools?
\end{enumerate}

We conducted a qualitative study with five Chinese legal practitioners to answer the research questions. We blended methods like think-aloud and interviews to understand legal practitioners' needs through multiple lenses (Sec.~\ref{sec:method}). 
Analyzing the multimodal data we collected (Sec.~\ref{sec:result}), we identified a pattern of the precedent search workflow: precedent search can be seen as an \textit{iterative sensemaking process}, the product of which impacts various stakeholders. Participants surfaced multiple challenges of today's precedent search tools, including a lack of support in search term development, limited reading assistance, and unpleasant context switching between tools. Meanwhile, users unanimously acknowledged the potential of AI but implied AIs should assist instead of automating the current workflow. Leveraging the qualitative findings, we articulate a potential prototype for a human-AI precedent search system (Sec.~\ref{sec:discussion}). We envision such a tool to 1) provide step-by-step AI assistance corresponding to the precedent search workflow that guarantees full user control and 2) leverage sensemaking techniques such as semantic zoom~\cite{bederson1994pad} to provide a seamless interface for users to navigate the precedent information space.

In summary, our study presents one of the first efforts at understanding how to design AI legal research tools through the lens of legal practitioners,  using precedent search in China as the use case. We subsequently contribute a design vision derived from our qualitative study to introduce further discussion around this topic.

%% file: sections/2-RelatedWork-short.tex
\section{Background and Related Work}
We begin our discussion by elaborating on the legal underpinnings of precedents (Sec.~\ref{subsec:rw-precedents}). Then, we discuss AI legal tools' current advancements and limitations (Sec.~\ref{subsec:rw-legalai-challenges}). Besides human-computer interaction (HCI) works, we often seek outside of HCI and include much literature from the legal and legal informatics domain for a comprehensive review.

\subsection{Precedents in Legal Practice}
\label{subsec:rw-precedents}

A precedent is a court decision that influences or decides subsequent cases with similar facts or legal issues~\cite{precedent-definition}.
Precedents are a cornerstone of the common law system, where they have a direct binding effect over judicial decisions~\cite{1920stare, douglas1949stare}. However, as a civil law jurisdiction, China does not recognize precedents as formal legal sources~\cite{lei2015rethinking, liu2021precedents}. %
Instead, blending elements of the civil law tradition with common law practices, the Chinese civil law system is supplemented by a case guidance system, requiring courts at all levels to reference guiding cases issued by the Supreme People's Court~\cite{zhang2017pushing, wang2019decoding, sun2019likecase}. 
Meanwhile, Judges from lower courts are evaluated by the rates of amendment and reversal of cases by higher courts, which strongly incentivizes them to follow precedents.
Therefore, while precedents in China do not have the same authority as common law countries, they play a significant role in shaping judicial practices~\cite{barioni2022emergence}, signifying the importance of precedent search in Chinese law practice~\cite{gu2019spontaneous}.

\subsection{AI-supported Legal Research}
\label{subsec:rw-legalai-challenges}

The past decades have seen numerous technical contributions at the intersection of law and computation, especially in the field of legal Natural Language Processing (NLP) (see reviews~\cite{chen2024a,ariai2024natural}). Recently, seeing the potential of legal NLP models, industry practitioners (e.g., MetaLaw~\cite{metalaw}, PkuLaw~\cite{pkulaw-ai}, and VitalLaw AI~\cite{vitallawai}) have been actively integrating NLP models to assist legal research--all claiming their products can ``speed up'' legal research process.

Despite evidence revealing AI assistance could improve productivity and quality of legal research~\cite{choi2023lawyering, weber2024legalwriter}, many research suggested limited success in supporting real-world legal practice~\cite{trozze2024large, deroy2023how, harasta2024it, magesh2024hallucinationfree, jaffe2024generative}. Among all, one key contributing factor is that today's AI does not perform ``as well as'' humans--they display decent performance on benchmarks yet are far from satisfactory for ethical and trustworthy use by practitioners~\cite{wendel2019promise}. For instance, \citet{magesh2024hallucinationfree} reported an alarming 17\% to 33\% ``hallucination'' (the phenomenon of ``\textit{generating plausible yet nonfactual content}''~\cite{huang2024survey}) rates even with the latest ``hallucination free'' (see relevant product claim~\cite{wellen2024hallucinationfree}) large language model-powered tools. Hence, using such imperfect algorithms adds substantial risk in legal practice, raising new considerations such as AI's unauthorized law practice and the unclear liability for inaccurate advice~\cite{cheong2024ai}.

We situate our work as a first step towards a human-AI~\cite{horvitz1999principles, amershi2019guidelines} legal research tool. Though applying the human-AI collaboration paradigm to design tools for critical societal domains~\cite{chen2024a} cannot be construed as new (e.g.,~\cite{zhang2024rethinking}), our work extends this discussion to legal practitioners, a demographics whose unique requirements are less discussed in the HCI community.

%% file: sections/3-Method.tex
\section{Method}
\label{sec:method}

We conducted a qualitative study (\textit{N} = 5) to answer the three research questions mentioned in Sec.~\ref{sec:intro}.

\subsection{Participants}
We applied a snowball sampling technique and recruited five participants. Our paper's third author, a junior associate lawyer himself, reached out on behalf of our research team via social media. We chose this approach to ensure most of our participants were early career lawyers--the stage where they would frequently engage in precedent search tasks. Specifically, three out of the five participants (60\%) identified as female; four participants reported job titles, three reported as Junior Associate Lawyers, and one as Legal Assistant\footnote{In this paper, ``Junior Associate Lawyers'' refer to individuals who have passed the legal bar exam in China and are qualified to practice law. ``Legal Assistants'' refer to individuals who have not passed the bar exam (and thus cannot practice law) but have gained experience in non-practicing roles.}. Participants practice various legal disciplines (see Appendix Table~\ref{tab:Lawyer}), and all participants reside and practice in China.

\subsection{Procedure}

Our experiment consists of three stages: 

\underline{\textit{Stage 1. Informed consent and pre-screening.}} 
After participants expressed interest, we first fully debriefed the study's purpose and obtained their consent to participate. Then, we disseminated a pre-screening survey to understand their legal background. In the survey, participants optionally provide their demographics, areas of legal expertise, the tools they commonly use for precedent search, and their exposure to AI legal tools. Additionally, we document their consent to record the study and whether they could allow us to 1) record their demonstration of precedent search with one of their ongoing cases, 2) record a case we provided to them\footnote{We provide this option because cases professionals are currently working on might be confidential. If they opt out of option 1) and choose option 2), we would provide them with three sample cases spanning civil, criminal, and administrative law to tailor to the participant's expertise.}, or 3) listen to their description of the precedent search process only.

\underline{\textit{Stage 2. Observed think-aloud precedent search task.}}
In this stage, we observed each participant as they demonstrated how they conducted a precedent search. For participants who authorized screen recording, we asked them to share their screens so that we could see their interactions with the multiple systems they used. While they conduct the precedent search, we prompt them to think aloud and verbalize their mental process. We did not interrupt their workflow during the precedent search task--we took notes and clarified with the participants with retrospective questioning to follow up and clarify the rationale of their decisions and operations.

\underline{\textit{Stage 3. Semi-structured interview.}}
Stage 3 builds on Stage 2 and deepens the discussion to surface the current precedent search challenges and users' expectations for AI legal tools. Specifically, we asked them about the challenges of the precedent search process and their concerns and vision regarding the use of AI in legal practice (see interview script in Appendix~\ref{sec:Interview Script}).

All studies were conducted online through VooV\footnote{\url{https://voovmeeting.com/}} to accommodate the busy schedule of our participants. The study was conducted by multiple authors in our team; all were briefed on the study goal and piloted the experiment. 
The ethics of study design and procedure were reviewed. Participants remained anonymous throughout the study. All screen recordings and audio were recorded and stored locally to safeguard the data from unauthorized third parties.

\vspace{-0.5em}

\subsection{Analysis}
We conducted qualitative coding on the audio transcripts to derive insights from the qualitative data, using screen recordings and the researcher's notes as context. Initially, three coders (the first, fourth, and fifth author) independently coded distinct subsets of the materials inductively and created personalized codebooks. Then, the three coders merged their personalized codebooks to derive a codebook that covered all the existing observations. After this, each of the three coders uses the current codebook to deductively code a new material previously unseen by the coder, flagging new code when necessary. We continuously refined the codebook through iterative collective discussions until no new codes emerged. Ultimately, led by the first author, the three coders collaboratively revisited the codes to surface themes from the data. %

%% file: sections/4-Finding.tex
\begin{figure*}[t]
\centering
\includegraphics[width=1\linewidth]{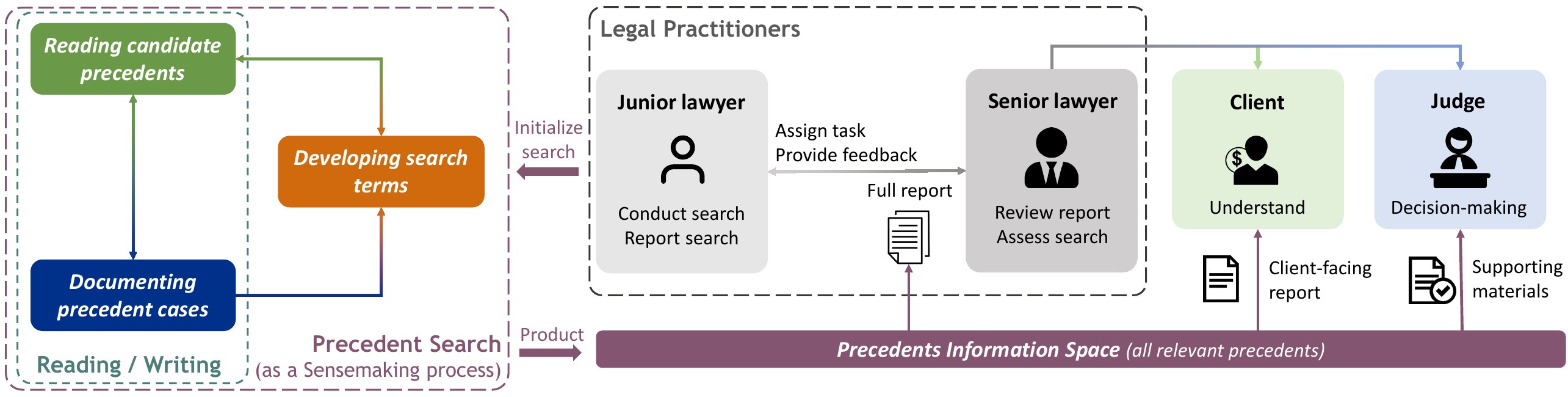} 
\caption{The workflow of precedent search. Precedent search consists of three iterative stages: developing search terms, reading candidate precedents, and documenting precedent cases. The product of precedent search also impacts multiple stakeholders, including junior lawyers, senior lawyers, clients, and judges.}
\label{fig:search process}
\end{figure*}

\section{Result}
\label{sec:result}
In this section, we report our findings following the three research questions we raised in Sec.\ref{sec:intro}. First, we report the workflow for precedent search (Sec.\ref{sec:workflow}). Then, we further document the insights about the challenges of precedent search (Sec.\ref{sec:challenge}) and practitioners' views of AI legal tools (Sec.\ref{sec:perception}).

\subsection{RQ1: Precedent Search Workflow}
\label{sec:workflow}

\begin{figure}[t]
  \centering
  \makeatletter
    \if@twocolumn
      \includegraphics[width=0.8\linewidth]{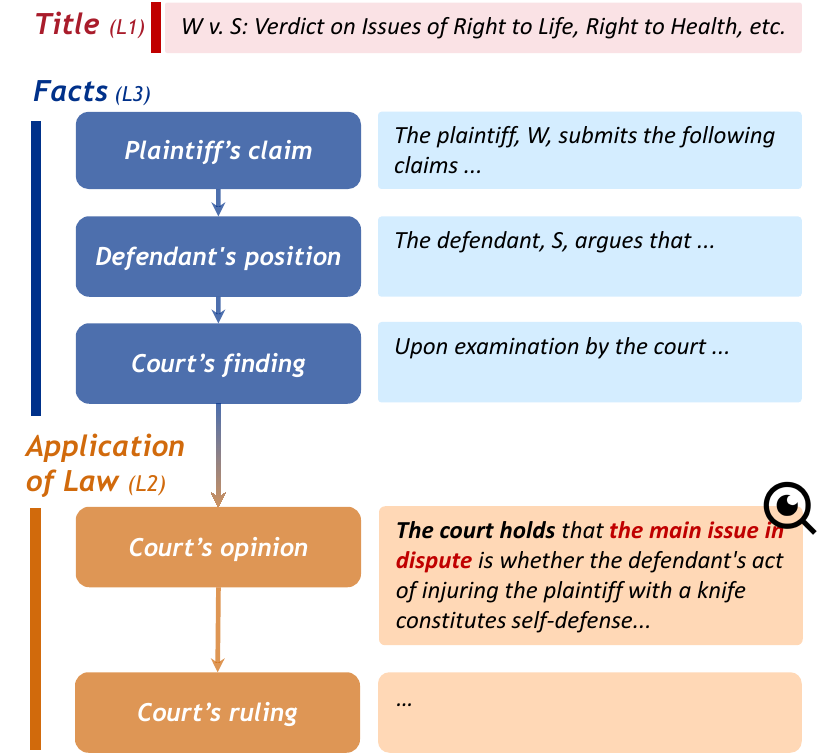}
    \else
      \includegraphics[width=0.4\linewidth]{figs/judgement.pdf}
    \fi
  \makeatother
  \caption{A typical structure of a Chinese Judgment Document: the document starts by stating the facts (blue sections) and ends with an explanation regarding the application of the law (orange sections)\protect\footnotemark.} %
  \label{fig:judgment}
\end{figure}

\subsubsection{Precedent Search Workflow}
We found that precedent search can be categorized into three phases (Fig.~\ref{fig:search process}): developing search terms, reading candidate precedents, and documenting precedent cases. Notably, those three phases do not follow a linear procedure--participants frequently switch between phases to progress the precedent search process. For instance, reading and documenting precedent cases can be a concurrent process as participants understand the candidate precedents. Inspired by existing precedents, new search terms emerge as they jump to the search term developing stage. We resonate this process with the sensemaking process~\cite{pirolli2005sensemaking}, where legal practitioners iteratively gather and synthesize information to construct meaning from a corpus of candidate precedents. In the following, we provide a detailed description of participants' representative actions in each phase.

\underline{\textit{Developing relevant search terms.}}
At the inception of the precedent search process, all participants began deriving initial search terms with legal knowledge, which they used to query legal databases. Iteratively, they consult relevant judicial interpretations and legal statutes (P3, P4) to further develop search terms or experiment with different search terms in multiple searches (P1, P2, P3, P4). This phase maps to the ``foraging loop'' in the sensemaking process.

\underline{\textit{Reading candidate precedents.}}
\label{para:reading-precedents}
After obtaining the search results, participants skim candidate precedents to select candidate precedents. Participants often apply a top-down approach to develop candidate precedents--initially, they relied solely on descriptive information (e.g., title) to determine the relevance of the document (\textbf{L1}). Accessing a specific candidate precedent, participants would first jump to a critical portion of the document related to the practice of law (\textbf{L2}) and then proceed to the factual portion of the document (\textbf{L3}) (see an illustration in Fig.~\ref{fig:judgment}). Participants considered ``issues in dispute'', typically in the ``court's opinion'' portion, a key indicator of relevance. For instance, P4 commented:\textit{``Then [after obtaining search results], we'll directly focus on the court's opinion and the issues in dispute.''} P3 further explained this top-down approach in reading candidate precedents: \textit{``... the first step is always to look at the court's opinion. Then, if it's closely related to the legal issue we're trying to solve, we'll continue by looking at the factual sections.''} This phase bridges the transition between the ``foraging loop'' and \footnotetext{Content extracted from \textit{W v. S}:\\ \scriptsize \url{https://www.pkulaw.com/pfnl/08df102e7c10f2063a0be7f9584fccac1be2ddaac92f35b9bdfb.html?}} the ``sensemaking loop'' in the sensemaking process.

\underline{\textit{Documenting precedent cases.}}
When participants find a precedent, they document and summarize it for future use. Participants reported key information they would typically document, including precedent number, court name, facts, judicial reasoning, and judgment. On top of information for individual precedents, participants would refine and summarize their findings to compile a precedent search report. P1 contextualized the organization of this documentation process: \textit{``[The report needs] relevant laws and regulations, the basic facts of some highly related cases, the rulings, and then the legal advice for this case.''} This phase maps to the ``sensemaking loop'' in the sensemaking process.

\subsubsection{Stakeholders}
Besides the precedent search workflow, we also revealed that precedent search is not a ``one-man job'': the product of this process actually impacts many stakeholders. Specifically, we found four entities in this process: junior lawyers, senior lawyers, clients, and judges. Junior lawyers directly contribute to precedent search, evidenced by a comment from P5:\textit{``It's usually the junior lawyer who does it (precedent search).''} The precedent search report compiled by junior lawyers is subsequently used by multiple parties. For instance, senior lawyers review these reports to 1) assess the quality of the precedent search and 2) make decisions about how to proceed with the case in hand: 
\textit{``They (senior lawyers) will review it to decide whether to accept it, figure out what parts they're interested in, and mark them. - P5''} Meanwhile, the client would also receive a version of the report that provides them with an expectation of the potential outcomes of their cases: \textit{``... for different clients, we selectively choose cases tailored to them. - P3''} Another stakeholder is the judges, who sometimes receive a report as part of the legal submission. This version of the report is also different from all others, as P2 said: \textit{``If it's for the judge, then the report actually needs to be a bit simpler.''} 

\subsection{RQ2: Challenges in Precedent Search}
\label{sec:challenge}

\subsubsection{Lack of support in search term development}
Developing search terms is not merely a process of summarizing the facts of a case; it is a hard mental process that involves identifying key legal terms, charges, and issues of dispute: \textit{``In actual case handling, the points that lawyers focus on are not just a simple summary of the case facts. - P2''} The challenge of developing search term are two-fold: 1) legal practitioners need advanced skills and undergo hard mental process to develop alternative search terms, and 2) documents use different wording to express the same non-legal concepts, which cannot be captured with word matching. P3 provided an example scenario:
\begin{quote}
\textit{``The challenge is that courts use different terms ... for example, in a case related to `year-end bonus 年终奖,' some other judges might use `annual performance bonus 年度绩效奖金.' In our view, both terms represent the same issue. However, the search results might be different, and we have to run two separate searches. This leads to some overlap [in search results and] increases our workload.''} - P3
\end{quote}
Though current systems might support fuzzy search, the search algorithm does not support a semantic understanding of terms, leaving this process solely manual to legal practitioners.

\subsubsection{Limited reading assistance}
Though participants reported prioritization strategies in reading precedents (from L1 to L3, see Sec.~\ref{para:reading-precedents}), some documents remain lengthy. P3 described one such scenario: \textit{``Sometimes, in larger commercial contracts or investment-related cases, there is a lot of data. As a result, both the case details section and the court's opinion are often excessively lengthy.''} Although today's legal search tools normally distinguish between L1 and L2/3 information, they generally provide limited assistance in helping practitioners understand L2 and L3 information.  

\subsubsection{Unpleasant context switching between tools}
Another phenomenon we observed was the extensive context-switching actions. For example, when P2 tried to record one precedent, she opened a text editor and pasted the relevant content from the browser. Surprisingly, she switched seven times between the text editor and the browser only to refine this note. This example signifies how a seemingly simple process in precedent search can become tedious.

\subsection{RQ3: Perceptions and Expectations for AI Legal Tools}
\label{sec:perception}
\subsubsection{Perception of AI legal tools}
Overall, the general sentiment of participants towards AI legal tools is positive, which aligned with related findings~\cite{WoltersKluwer2024report}. However, despite the enthusiasm, participants raised several concerns regarding AI legal tools. Participants pointed out they are concerned about AI tools omitting critical information and generating inaccurate or fabricated materials. For example, describing a specific legal point, P2 commented: \textit{``If (AI) summarize it directly, I think it's very likely that a point would be overlooked.''} Moreover, participants perceive AI tools cannot conduct complex legal reasoning, taking a conservative stand in automating tasks requiring deep legal insight and expertise. For instance, P1 thought AIs could not identify legal issues ``end-to-end'': \textit{``When it comes to identifying legal issues (from scratch), I feel like it's still pretty hard to achieve accurate recognition right now.''}

\subsubsection{Expectation of AI legal tools}
Participants generally believed that AI legal tools should assist lower-level tasks instead of automating the whole process. For example, P1 believes AI can handle \textit{``basic and simple legal issues''}, while P5 argued AI can be automate ``\textit{less serious}'' works and \textit{``then go back and double-check.''}

We also revealed that participants' expectations strongly correlate with the workflow processes. For example, P2 and P4 envisioned an AI-assisted search term development process, where AI goes beyond summarizing the facts and generates a corpus of search terms for them to choose from. Note that both participants suggested that the search process should still be handled by themselves instead of automating it. Participants also suggested a vision of a faceted fact summarization tool to expedite the reading process. P3 described this vision: \textit{``If you can, like, just pull out the facts that are, you know, related to this (the legal issue to be addressed), and the legal views that are just about this, I think it'd really help a lot.''} While some expected help from summarization tools, others implied the importance of the original text. For instance, while P2 agrees with the value of summaries, she pointed out that those texts should not be trusted when dealing with specific issues: \textit{``I guess I'd treat it (the summary) as something I need to use at first. But once it gets to a specific focus and how to judge it, I probably wouldn't rely on it.''}

Overall, P4 provided a representative summary of legal practitioner's expectations of AI tools:
\begin{quote}
\textit{``Considering its (AI's) limitations, I think it's still more reliable to leave the [precedent search] work
to humans. It's probably better to just let it (AI) do the organizing and summarizing work.''} - P4
\end{quote}

%% file: sections/5-Discussion.tex
\section{Discussion}
\label{sec:discussion}

\subsection{Envisioning the Human-AI Precedent Search Tool}
Our findings suggested two major insights that we carry to inform the design of a human-AI precedent search tool. Those insights include 1) we can view precedent search as a sensemaking process, 2) legal practitioners lack lower-level task support, and they expect AIs to support those tasks while maintaining full control themselves for higher-level legal reasoning.

We propose a prototype (Fig.~\ref{fig:system}) to reflect our findings. Specifically, we leveraged the semantic zoom~\cite{bederson1994pad} technique to support the sensemaking process of precedent search. Additionally, we designed the system following the guidelines of human-AI interaction~\cite{horvitz1999principles, amershi2019guidelines} to ensure human control. The three key features of our proposed design include:

\begin{figure*}
    \centering
    \includegraphics[width=1\linewidth]{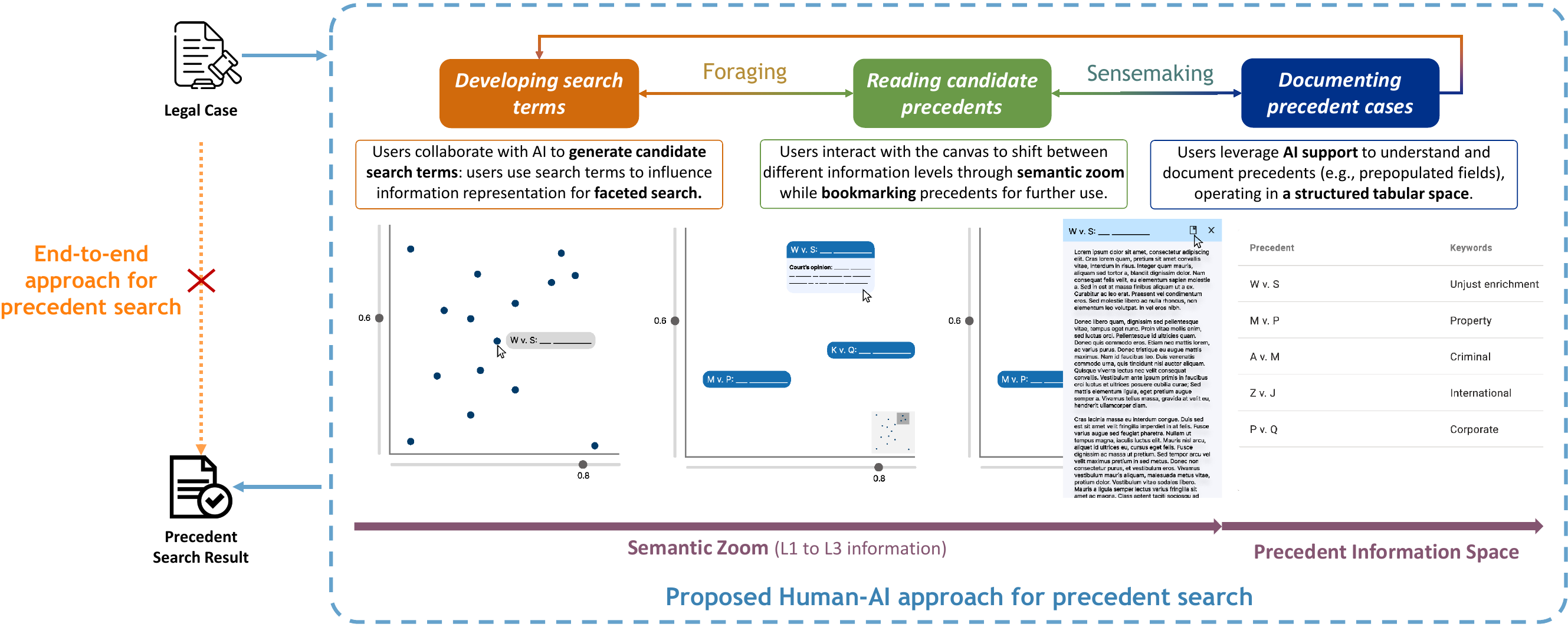}
    \caption{The schematics of our proposed Human-AI precedent search tool. We provide step-by-step assistance instead of an end-to-end approach while we model precedent search as a sensemaking process~\cite{pirolli2005sensemaking}.}
    \label{fig:system}
\end{figure*}

\underline{\textit{Visualized Faceted Search.}} 
Participants have pointed out that one key challenge is the lack of support in search term development (Sec.~\ref{sec:challenge}). On this challenge, we envision that AI could collaboratively recommend or generate alternative search terms~\cite{lyu2019synonyms}. The user-generated or system-generated search terms could then be assigned to one of the two axes to initialize a faceted search. Users could combine words with similar meanings or closely related legal terms, adding flexibility in the foraging stage to make sense of the precedent space. Users could also customize the similarity threshold on each facet to control the size of the visualized space. 

\underline{\textit{Semantic Zoom.}} 
In addition to the search term development, participants highlighted a lack of reading support and excessive context switching. By modeling the precedent search as a sensemaking process, we could use the semantic zoom technique~\cite{bederson1994pad} to facilitate precedent understanding through assisted reading. Specifically, following the three levels of information in a precedent (see Sec.~\ref{sec:workflow}), we defined three granularity of information representation. We envision users can seamlessly switch between the information levels and leverage related AI reading functions (e.g., summarization, citation to search terms, etc.) to collaborate with AI. We also envision a bookmark function bridging the transition between unstructured exploration and structured documentation. Specifically, all bookmarked precedents would be presented in a tabular form for further processing. Additionally, AI tools could help in this stage by suggesting different versions of precedent reports to accommodate the needs of multiple stakeholders. 

\underline{\textit{Step-by-step Assistance.}} 
Aligning to user expectations (Sec.~\ref{sec:perception}), our prototype provides step-by-step assistance to give users full control. We designed the type of assistance to be low-level: legal inferences remained in full control of the users. We expect this design to better align with the familiar workflow of legal practitioners.

\subsection{Limitations and Future Work}
In this section, we describe the limitations of our study and envision future research opportunities:

\underline{Expanding participant demographics.} Our study's recruitment criterion dictated a very niche demographic: junior lawyers who practice precedent search tasks in China. Although we believe our choice is validated for our research question, other angles (e.g., other stakeholders in precedent search, practitioners from other legal systems) exist. Future work could expand towards a more comprehensive viewpoint and understanding of how AI legal tools might affect their workflow and affordances.

\underline{Going beyond precedent search.} To limit the scope of the study, we chose precedent search as a specific task to probe the needs of legal practitioners for a human-AI legal research system. Though generic findings (e.g., step-by-step guidance) might generalize to other legal tasks, we acknowledge that findings specific to precedent search (e.g., develop search terms) might not generalize to other legal research tasks. Future work should explore a more generic guideline regarding the design of human-AI legal research systems.

\underline{Generalizing over legal systems.} Another limitation stemmed from our study's context: we based our study on Chinese law, a variant of the civil law system. In the Chinese legal system, precedents serve only as supplemental material in legal reasoning, with the court decision still based on the statutory law. Meanwhile, in common law systems, precedents play a more significant role than the Chinese law systems, which might impact how practitioners approach precedent searches. For example, practitioners in common law systems might require further support in legal reasoning and writing based on precedents. Future work should investigate whether, if so, how our findings generalize to other legal systems.

%% file: sections/8-Conclusion.tex
\section{Conclusion}
By understanding the needs of legal practitioners for precedent search tasks, our study serves as a first step for designing human-AI interfaces to address the current challenges of AI adoption in the legal domain. In addition to the qualitative findings, we also make an initial contribution to the design of AI precedent search systems that are tailored to the user's needs. We hope our contribution could spark further discussion in the HCI community about intelligent legal interfaces towards a goal to responsibly use AI to support the critical societal domain of law.

%% file: sections/9-appendix.tex
\newpage

\section{Participant Demographics}

Table~\ref{tab:Lawyer} documented the demographics of our participants.

\begin{table}[t]
\centering
\small
\caption{Demographics of Participants.}
\begin{tabular}{clll}
\toprule
\textbf{ID} & \textbf{Gender} & \textbf{Area of Expertise} & \textbf{Job Title}  \\ \midrule
P1 & Male & Entertainment Law & Junior Associate Lawyer  \\
P2 & Female & International Law & -  \\
P3 & Female & Labor Law & Junior Associate Lawyer   \\
P4 & Male & Financial Law & Legal Assistant \\
P5 & Female & International Law & Junior Associate Lawyer  \\ \bottomrule
\end{tabular}
\label{tab:Lawyer}
\Description{This table shows the demographic information of five research participants, including gender, area of expertise, and job title. Three participants (60\%) are female. Participants' areas of legal expertise include Entertainment Law, International Law, Labor Law, and Financial Law. The participants' job titles include Junior Associate Lawyer (3) and Legal Assistant (1). Participant 1 is a male Junior Associate Lawyer specializing in Entertainment Law. Participant 2 is a female with expertise in International Law. She chose not to report her job title. Participant 3 is a female Junior Associate Lawyer specializing in Labor Law. Participant 4 is a male Legal Assistant specializing in Financial Law. Participant 5 is a female Junior Associate Lawyer with expertise in International Law.}
\end{table}

\section{Interview Script}
\label{sec:Interview Script}

We provide the scaffold of the interview questions we conducted at the end of each study.

\begin{enumerate}
    \item[\textbf{Q1}] \textbf{Challenges in precedent search}: What difficulties have you encountered during precedent search?
    \item[\textbf{Q2}] \textbf{Precedent search workflow}: What are your habits when reading candidate precedents, and what aspects do you primarily focus on? What elements are typically included in your precedent search report?
    \item[\textbf{Q3}] \textbf{Perceptions of AI legal tools}: Have you used an AI-based precedent search system? What do you think are the current problems with this type of product? What is your stance towards integrating AI into your work?
    \item[\textbf{Q4}] \textbf{Expectation of AI legal tools}: In which aspects do you think AI could assist precedent search? If you are to envision an AI legal assistance tool, what functionalities do you think are needed to support precedent search?
    \item[\textbf{Q5}] \textbf{Closing question}: Is there anything else you want to share with us or any questions about our interview?
\end{enumerate}